%% file: llp_main.tex
\documentclass[conference]{IEEEtran}
\IEEEoverridecommandlockouts
\usepackage{cite}
\usepackage{amsmath,amssymb,amsfonts}
\usepackage{algorithmic}
\usepackage{graphicx}
\usepackage{textcomp}
\usepackage{xcolor}
\usepackage{svg}
\usepackage{afterpage}
\usepackage{booktabs}
\usepackage{float}

\def\BibTeX{{\rm B\kern-.05em{\sc i\kern-.025em b}\kern-.08em
    T\kern-.1667em\lower.7ex\hbox{E}\kern-.125emX}}
\begin{document}

\title{Accelerating Local Laplacian Filters on FPGAs}


\author{\IEEEauthorblockN{Shashwat Khandelwal\IEEEauthorrefmark{1},
Ziaul Choudhury\IEEEauthorrefmark{1}, Shashwat Shrivastava\IEEEauthorrefmark{1} and
Suresh Purini\IEEEauthorrefmark{2}}
\IEEEauthorblockA{Computer Systems Group,
International Institute of Information Technology\\
Hyderabad, India\\
Email: \IEEEauthorrefmark{1}\{shashwat.khandelwal, ziaul.c, shashwat.shrivastava\}@research.iiit.ac.in
\IEEEauthorrefmark{2}suresh.purini@iiit.ac.in}}

\maketitle

\input{abstract.tex}

\input{introduction}
\input{background}

\input{architecture}

\input{experiments}

\input{conclusions}

\bibliography{references}
\bibliographystyle{ieeetr}

\end{document}

%% file: abstract.tex
\begin{abstract}
Images when processed using various enhancement techniques often lead to edge degradation and other unwanted artifacts such as halos. These artifacts pose a major problem for photographic applications where they can denude the quality of an image. There is a plethora of edge-aware techniques proposed in the field of image processing. However, these require the application of complex optimization or post-processing methods. Local Laplacian Filtering is an edge-aware image processing technique that involves the construction of simple  Gaussian and Laplacian pyramids. This technique can be successfully applied for detail smoothing, detail enhancement, tone mapping and inverse tone mapping of an image while keeping it artifact-free. The problem though with this approach is that it is computationally expensive. Hence, parallelization schemes using multi-core CPUs and GPUs have been proposed. As is well known, they are not power-efficient, and a well-designed hardware architecture on an FPGA can do better on the performance per watt metric. In this paper, we propose a hardware accelerator, which exploits fully the available parallelism in the Local Laplacian Filtering algorithm, while minimizing the utilization of on-chip FPGA resources. On Virtex-7 FPGA, we obtain a 7.5x speed-up to process a 1 MB image when compared to an optimized baseline CPU implementation. To the best of our knowledge, we are not aware of any other hardware accelerators proposed in the research literature for the Local Laplacian Filtering problem. 
\end{abstract}
 

%% file: introduction.tex
\section{Introduction} 
Laplacian pyramids~\cite{burt1983laplacian} are multi-scale representations of images that are widely used in image processing as they are easy to build. But these pyramids are considered ill-suited for applications which require edge-aware processing as they are built with spatially invariant Gaussian kernels. Since they don't take edge discontinuities into account, it leads to edge degradation and the introduction of other artifacts such as halos~\cite{tsutsui2012halo}. Many techniques for edge-aware image processing, such as anisotropic diffusion~\cite{aubert2006mathematical, perona1990scale}, neighbourhood filters~\cite{tomasi1998bilateral}, edge-aware wavelets~\cite{fattal2009edge} and edge-preserving optimizations~\cite{bhat2010gradientshop, farbman2008edge} have been proposed in the literature. Though these techniques have been quite successful in producing artifact-free images they often do it with the use of complex non-linear methods. Another technique to solve the problem of edge artifacts is Local Laplacian Filtering~\cite{paris2011local} which is an edge-aware technique based on simple Laplacian pyramids for detail enhancement, detail smoothing, tone mapping, and inverse tone mapping.  

In this technique for every pixel in the Gaussian pyramid of the input image, there is a small section in the original image which is passed through a remapping function. A Laplacian pyramid for this remapped image is constructed and the pixel corresponding to the Gaussian pixel which we are processing is picked up and updated in the output Laplacian pyramid. 
We present the detailed algorithm in Section~\ref{sec:background}. This technique is computationally very expensive as for each pixel in the Gaussian pyramid a Laplacian pyramid is constructed. Hence, the original paper on Local Laplacian Filtering~\cite{paris2011local} gives a multi-core implementation for the same. Aubry et al.~\cite{aubry2014fast} proposed a parallelization approach using GPUs. However, as it is well-known, FPGAs fare way better than multi-core CPUs and GPUs on the performance per watt metric. So it is natural to consider accelerating Local Laplacian Filters on FPGAs. To the best of our knowledge, there is no prior work on this problem. However, there is prior work on FPGA realization of other edge-aware techniques such as bilateral filters~\cite{dabhade2017reconfigurable, gabiger09bilateral}. 

While there is a lot of parallelism available in the Local Laplacian Filter computation, it is very difficult to exploit the same, due to the complexity of the underlying parallelism structure. This is especially so in designing hardware structures. In this paper, we propose an FPGA based hardware architecture that fully exploits all forms of available parallelism, albeit complex. Further, we adapted the Gaussian filter slightly, which enabled us to design a high throughput convolution engine which is both pipelined and data parallel. The convolution engine requires no DSP blocks on the FPGA and uses minimal LUTs. The remap function computation is turned into a table lookup operation thereby saving substantial hardware resources. This also helped in eliminating associated computational latency. We verified the accuracy of our implementation by comparing it with the original implementation using the PSNR metric. Finally, through a set of experiments, we have found our implementation to be 7.5x times faster than the original CPU version.  

The layout of the rest of the paper is as follows. Section~\ref{sec:background} provides the detailed algorithm for Local Laplacian Filtering; Section~\ref{sec:architecture} presents our novel parallel architecture; Section~\ref{sec:experiments} contain the experimental results; and finally we conclude in Section~\ref{sec:conclusions}.

%% file: background.tex
\section{Background}\label{sec:background}
In this section, we present the necessary background on Local Laplacian Filtering algorithm which is essential for understanding the proposed accelerator architecture in Section~\ref{sec:architecture}.
\subsection{Gaussian and Laplacian Pyramids}
The Gaussian pyramid for an input image $I$ is a set of images $\{G_l~|~0\leq l \leq n\}$ such that $G_0 = I$ and $G_{l+1} = downsample(G_l)$. Downsampling an image involves application of Gaussian filter followed by a sub-sampling procedure. The dimensions of $G_{l+1}$ are half that of $G_l$. Thus, the image sequence in the Gaussian pyramid is an increasingly low resolution representation of the original image $I$. Associated with a Gaussian pyramid $\{G_l\}$, we can construct a Laplacian pyramid $\{L_l\}$, wherein $L_l = G_l - upsample(G_{l+1})$. The upsampling operation doubles the dimensions of the image $G_{l+1}$ using a smoothing operation. Given the Gaussian and Laplacian pyramids, we can collapse the Laplacian pyramid from the top by iteratively applying the operation $G_l = L_l + upsample(G_{l+1})$ until $l=0$. 
\subsection{Local Laplacian Filtering}\label{sec:local}
Image processing using basic Gaussian and Laplacian pyramids can lead to edge artifacts in applications such as detail enhancement and tone mapping. The Local Laplacian Filtering algorithm described in this section shows how its edge-aware approach circumvents these problems.

The Gaussian and Laplacian pyramids for an image $J$ are denoted by $G[J]$ and $L[J]$ respectively. The image $J$ can be the input image or a sub-image of the input image. $G_l[J]$ denotes level $l$ of the Gaussian pyramid and $G_l[J](x,y)$ denotes the pixel at $(x,y)$ position within the level $l$. 
We use analogous notation $L_l[J]$ and $L_l[J](x, y)$ for Laplacian pyramid.
The first step in Local Laplacian Filtering is the construction of the Gaussian pyramid $G[I]$. For every pixel $g = G_l[I](x,y)$ of the Gaussian pyramid, a sub-image $R$ in the input image is identified. This sub-image is passed through a remapping function. The remapping functions categorizes each pixel $i$ in the sub-image as an edge or a detail. If $|i-g| \leq \sigma$, where $\sigma$ is a user-defined parameter, then the pixel is treated as a fine-scale detail, which is modified by the detail remapping function $r_d$ as follows.
\begin{equation}
  r_d(i,g,\sigma) = g + sign(i-g)\sigma f_d(|i-g|/\sigma)  
\end{equation}
The function $f_d$ 
is a detail manipulation function which performs smoothing (if $\alpha > 1$)  or enhancement (if $0 < \alpha < 1$) operations, where $\alpha$ is an user-defined parameter. 
On the other hand, a pixel $i$ is treated as an edge if $|i-g|> \sigma$.  Then the pixel is modified by applying the following edge-remapping function $r_e$.  
\begin{equation}
r_e(i,g,\sigma) = g + sign(i-g)(f_e(|i-g|-\sigma)+ \sigma)    
\end{equation}
The edge-aware tone mapping function $f_e$ 
performs tone mapping, (if $0\leq \beta < 1)$, and inverse tone mapping, if $\beta > 1$, where $\beta$ is a user-defined parameter.

Let $\Tilde{R}$ denote the remapped image. Then the Gaussian pyramid $G[\Tilde{R}]$ and the Laplacian pyramid $L[\Tilde{R}]$ of the remapped image are constructed. Recall that we are trying to compute the Laplacian coefficient $L_l[I](x,y)$ corresponding to the pixel $g=G_l[I](x,y)$ in an edge-aware fashion. Hence, the Laplacian coefficient $L_l[\Tilde{R}](x,y)$ is used in the place of $L_l[I](x,y)$. This process is followed for every pixel $G_l[I](x,y)$ to completely generate the edge-aware output Laplacian pyramid $\Tilde{L}[I]$. Note that while $L[I]$ denotes the regular Laplacian pyramid, $\Tilde{L}[I]$ denotes the Laplacian pyramid constructed using the Local filtering method. 
The generated output Laplacian pyramid can be collapsed to construct the enhanced input image without introducing any edge artifacts.

%% file: architecture.tex
\begin{figure*}[t]
     \centering
     \includegraphics[width=\textwidth]{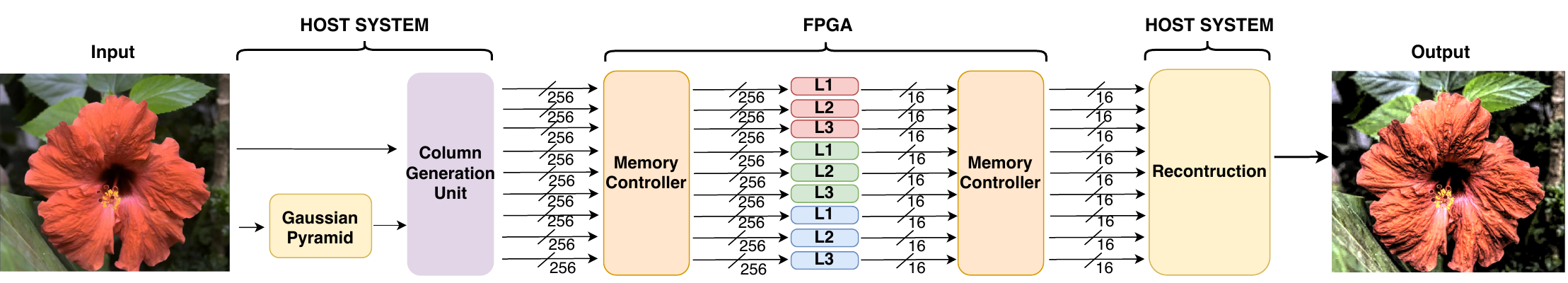}
     \caption{Overview of the hardware architecture depicting the flow of data from the host to FPGA and then back to the host. It shows the example of a detail enhancement operation.}
     \label{fig:cpufpga}
\end{figure*}                    
\section{Proposed Architecture}\label{sec:architecture}
This section describes our hardware architecture in detail. The Gaussian pyramid associated with the input image $G[I]$ is constructed on the host. The construction of the Laplacian pyramid $\Tilde{L}[I]$ which is the most computationally expensive happens on the FPGA. Finally, image reconstruction from the Laplacian pyramid is done on the host system. The hardware for Laplacian pyramid construction on the FPGA broadly consists of 9 processing units. All the 3 RGB channels in an image are processed in parallel. For each channel, three levels of output Laplacian pyramid $\Tilde{L}_1$, $\Tilde{L}_2$ and $\Tilde{L}_3$ are constructed in parallel. Each level $\Tilde{L}_i$, for $i=0, 1, 2$ is computed by a Level Processing Unit (LPU) $\Tilde{L}_i$. Figure~\ref{fig:cpufpga} shows an overview of the hardware architecture and the data flow in the heterogeneous system. Each LPU consists of a remap unit, a convolution engine, a downsampling unit and an upsampling unit as shown in Figure~\ref{fig:lpu}.  Data is fed from the host to the nine processing units independently. This is done by creating nine separate input streams. The output data from the nine processing units is transferred to the host from the FPGA using nine independent output streams. 

\subsection{Remap Unit}
\begin{figure}[t]
    \centering
    \includegraphics[width=8cm, height=4.5cm]{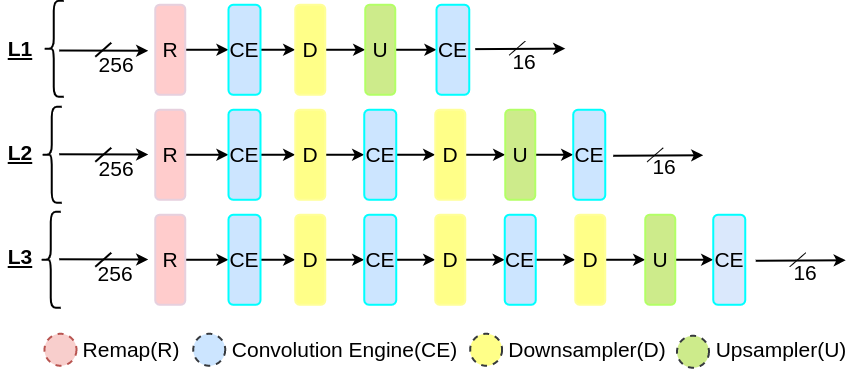}
    \caption{LPUs $\Tilde{L}_1$, $\Tilde{L}_2$ and $\Tilde{L}_3$, each comprising of a remap unit, a convolution engine, a downsampling unit and an upsampling unit.}
    \label{fig:lpu}
\end{figure}

From Figure~\ref{fig:lpu}, we can notice that the first stage in any LPU is a Remap Unit. Recall from Section~\ref{sec:local}, the remap function treats a pixel $i$ as a detail or an edge depending on whether $|i-g|\leq \sigma$. If it is a detail, then the detail function $r_d$ is applied, otherwise, the edge function $r_e$ is used. These remap functions involve addition, multiplication and exponentiation operations. Exponential functions can be approximated, but are still difficult to handle in hardware.  
 
 The remap functions $r_d(i, g, \sigma)$ and $r_e(i, g, \sigma)$ invoke the functions $f_d$ and $f_e$ respectively. These functions depend on the user-defined parameter $\alpha$, $\beta$ and $\sigma$; and on the pixel difference value $|i-g|$. The user-defined parameters remain constant once fixed. Since the pixels $i, g \in [0,255]$, the absolute value of their difference $|i-g| \in [0, 255]$. Based on this observation, for a given $\alpha$, $\beta$, $\sigma$ and $0\leq |i-g| \leq 255$, we pre-compute the sub-expressions $\sigma f_d(|i-g|/\sigma)$ and $f_e(|i-g|-\sigma)+ \sigma$, necessary for computing the functions $r_d$ and $r_e$ respectively. These pre-computed values are stored in a look-up table on FPGA and is indexed by $|i-g|$ value.  Thus complex computations at runtime are completely eliminated saving on computational latency and FPGA hardware resources. 
\subsection{Convolution Engine}
After a remap function is applied to a sub-image associated with a pixel, we apply Gaussian filter followed by a downsampling operation. Depending on whether we are constructing $\Tilde{L}_1$, $\Tilde{L}_2$ or $\Tilde{L}_3$ level, we iterate this filter followed by downsampling operation that many times. For example, in Figure~\ref{fig:lpu}, we can notice that in $\Tilde{L}_2$ LPU, that iteration happens for 2 times.\par
Applying Gaussian filter on an input image is a two-dimensional convolution operation. As convolution operations involve Multiply-and-Accumulate operations (MACs), they are usually realized using the FPGA DSP blocks. Hence the number of convolutions that can be performed in parallel is constrained by the DSP availability. We circumvent this problem by modifying the convolution operation so that it can be computed using a simple shift-and-add operation instead of a costly MAC. The shift-and-add operation can be realized using only LUTs, which are present in a relatively higher percentage on the FPGAs. Consequently, our hardware at its peak is able to sustain 783 3x3 convolutions per cycle without using any DSP units.\par 
The Convolution Engine (CE) in Figure~\ref{fig:lpu} performs convolutions on an input image in a streaming fashion exploiting both pipelined and data parallelism. We approximated the original Gaussian filter proposed in~\cite{burt1983laplacian} with the following 3x3 filter $\widehat{G}$ which is more amenable for our shift-and-accumulate way of computation.
\begin{align*}
\widehat{G} = 
\begin{bmatrix}
\frac{1}{2^{\alpha}} & \frac{1}{2^{\alpha-1}} & \frac{1}{2^{\alpha}} \\
\frac{1}{2^{\alpha-1}} & \frac{1}{2^{\alpha-2}} & \frac{1}{2^{\alpha-1}} \\
\frac{1}{2^{\alpha}} & \frac{1}{2^{\alpha-1}} & \frac{1}{2^{\alpha}}
\end{bmatrix}
\end{align*}
The parameter $\alpha$ is a scale factor and we have set it to 4 in our experimental results. Note that the second column of the filter matrix is obtained by multiplying the first column by 2 and the third column is obtained by dividing the second column by 2. We use this observation to design a 3-stage pipelined engine for convolution operations. Further, the convolution engine is designed to perform convolutions on $l$ rows of an image simultaneously. A 3x3 convolution on $l$ rows results in simultaneously rendering $l-2$ rows of the output filtered image in a streaming fashion.
\begin{figure}[t]
    \centering
    \includegraphics[width=8cm]{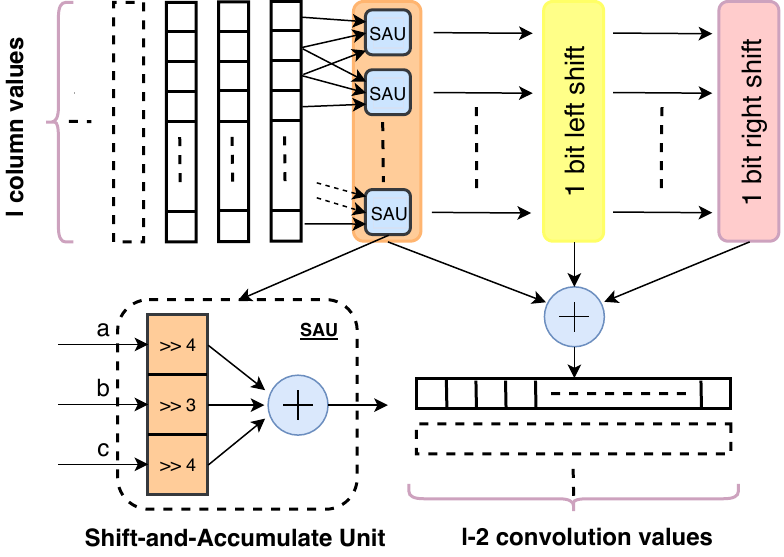}
    \caption{Architecture of the Convolution Engine which exploits both pipelined and data parallelism simultaneously.}
    \label{fig:ce}
\end{figure}
The first stage of the convolution engine takes a vector $X_0$ of size $l$.  and generates a vector $X_1$ of size $l-2$. The element $X_1[i]$ of the output vector is computed by the Shift-and-Accumulate Unit $SAU_i$ as follows.
\begin{align*}
    X_1[i] = X_0[i-1] >> 4 + X_0[i] >> 3 + X_0[i+1] >> 4
\end{align*}
Thus, the first stage contains an array of $l-2$ Shift-and-Accumulate units. The second stage of the pipeline takes the $l-2$ dimensional vector $B$ and generates another vector $C$ of same dimension as follows. 
\begin{align*}
    X_2[i] = X_1[i] << 1
\end{align*}
The third stage of the pipeline takes the $l-2$ dimensional vector $C$ and generates another vector $D$ of same dimension as follows.
\begin{align*}
    X_3[i] = X_2[i] >> 1
\end{align*}
After an initial latency of 2 clock cycles, once the pipeline becomes full, the first column of the filtered output image is nothing but the sum of the $l-2$ dimensional vectors $X_1$, $X_2$ and $X_3$ present at different stages in the pipeline. Thus, the convolution engine uses both pipelined parallelism and data parallelism across rows to maintain high throughput. Figure~\ref{fig:ce} depicts the architecture of the convolution engine described in this section. 

\subsection{Downsampling and Upsampling Units}
Each LPU also consists of downsampling and upsampling units. Downsampling is done by taking every alternate value from the input row/column and giving them as output. Upsampling is done by inserting zeroes in every alternate value in the input row/column and giving them as output. Note that the downsampling operation follows a filter operation in the Gaussian pyramid construction. And, upsampling is followed by a filter operation to construct the required level of the Laplacian pyramid.

\subsection{Summary of Parallelization Scheme}
\begin{figure}[t]
    \centering
    \includegraphics[width=8cm]{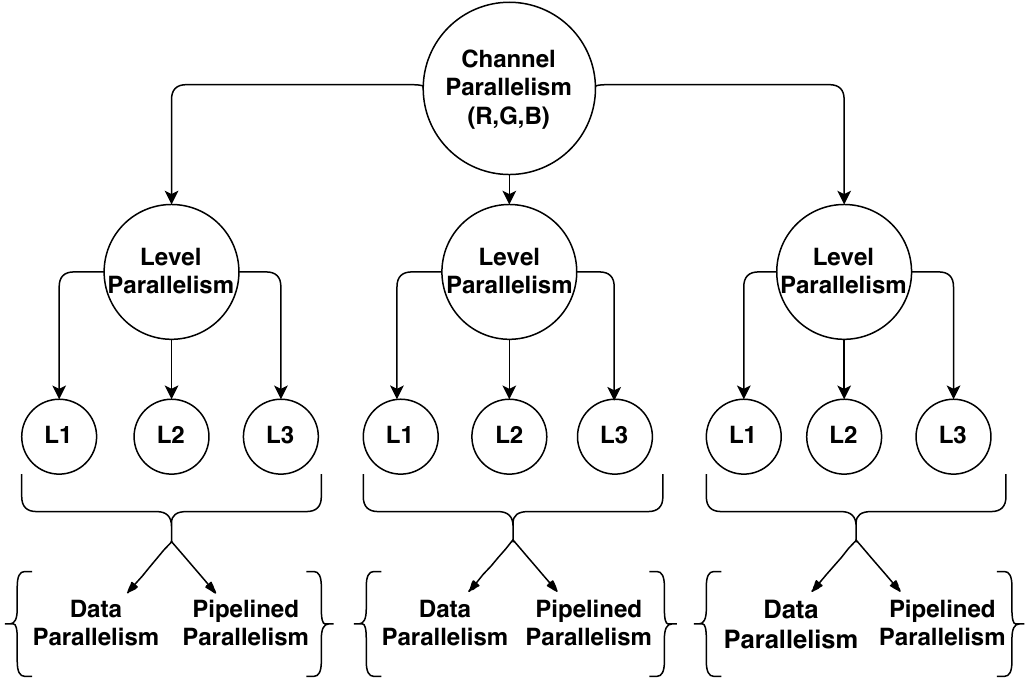}
    \caption{Summary of the parallelization scheme.}
    \label{fig:parallelism}
\end{figure}

We have planned the hardware architecture in a way to make use of all kinds of parallelisms which the algorithm has to offer. First, we make use of the channel parallelism, where we process the three RGB channels  of the image in parallel as shown in the Figure~\ref{fig:parallelism}. We then use level parallelism within a channel to compute the 3 levels $\Tilde{L}_1$, $\Tilde{L}_2$ and $\Tilde{L}_3$ of the output Laplacian pyramid simultaneously. Inside each level, we exploit pipelined parallelism to process each pixel of the output laplacian pyramid. We also use data parallelism, as we process entire columns of the sub-image in a single shot, while performing all the operations in a pipeline, rather than processing it pixel by pixel.


%% file: experiments.tex
\section{Experimental Results}\label{sec:experiments}
All the experiments are performed on a Virtex-7 FPGA connected to a 3 GHz Intel Core-i5 through a 3.5 GBPS PCIe-8x link. The host code is written in C++11 and compiled on a Ubuntu-16.4 system using GCC-5.4. Our hardware is synthesized at a clock frequency of 100 MHz. The three Level Processing Units in our accelerator are synthesized as independent IPs and stitched together in the whole system design. Depending on the availability of FPGA resources and memory bandwidth, it is possible to instantiate LPUs which are deeper and hence can handle higher layers of Laplacian pyramids.  

The original baseline implementation due to Paris et al.~\cite{paris2011local} uses a 5x5 Gaussian filter and constructs all the levels of the output Laplacian pyramid $\Tilde{L}[I]$. On a 2.2 GHz Intel Xeon CPU with 8 cores, for one megapixel image, they report that using a single thread the computation takes a minute. They optimize on time by reducing the number of levels in the intermediate Laplacian pyramids to 5. This is achieved by considering the sub-image at level $\max\{0, l-3\}$ while processing a pixel at level $l$. The remap function is applied on the coarse-grained sub-image rather than the sub-image at a full resolution. This reduces the size of the sub-image substantially as the level $l$ increases, thereby saving on computations. However, there could be a loss in accuracy. The claim though is that the loss in accuracy is visually indistinguishable from the full pyramid process with a PSNR ranging from 30 to 40 dB. The modified approach took 4 seconds while using 8 cores for the same one mega-pixel image.  

In this paper, we construct only three levels $\Tilde{L}_i$, $0 \leq i \leq 3$ of the output Laplacian pyramid using 3x3 modified Gaussian filter. The intermediate Laplacian pyramids are constructed by using the full resolution sub-image though. We compare our approach with that of the baseline on the latency and accuracy metrics. 
\subsection{Latency}
Table~\ref{tab:lputime} shows that our accelerator can process a 1 mega-pixel image in 534 milli-seconds (ms) as against 4 seconds time taken by the 8-core CPU implementation. This gives us 7.5x speedup over the baseline. 
It also shows the latency for various image sizes. 

For example, the last row of the table shows the time taken when the image size is one mega-pixel. As expected the latency associated with the level $\Tilde{L}_1$ is the highest when compared with the other levels as it computes the maximum number of pixels in the output Laplacian pyramid. The column labelled {\it sequential} denotes the time taken if the three levels are processed sequentially as against the time taken when they are processed in {\it parallel} which is the case in our architecture.
\renewcommand{\arraystretch}{1.2}
\begin{table}[ht]
\centering
\begin{tabular}{|c|c|c|c|c|c|}
\hline
\textbf{Size (MP)} & \multicolumn{5}{c|}{\textbf{Latency (ms)}}                                        \\ \hline
                  & \textbf{L1} & \textbf{L2} & \textbf{L3} & \textbf{Sequential} & \textbf{Parallel} \\ \hline
\hline
0.25               & 133         & 64          & 49          & 246                 & 133               \\ \hline
0.5                & 267         & 129         & 99          & 495                 & 267               \\ \hline
0.75               & 400         & 194         & 148         & 772                 & 400               \\ \hline
1                  & 534         & 259         & 198         & 991                 & 534               \\ \hline
\end{tabular}
\caption{Latency of the LPUs $\Tilde{L}_1$, $\Tilde{L}_2$ and $\Tilde{L}_3$ while processing images of different sizes. }
\label{tab:lputime}
\end{table}
\subsection{Accuracy}
We compared our implementation against the baseline CPU implementation wherein the intermediate Laplacian pyramids are constructed from the full resolution sub-image as against the downsampled sub-image. Note that the latencies reported in the previous sub-section is for the downsampled sub-image approach. 
Table~\ref{tab:psnr} shows the PSNR values for various values of the user-defined parameters $\alpha$, $\beta$ and $\sigma$. The PSNR values range between 30 to 50 dB with most values close to or above 40 dB.

\begin{table}[t]
\begin{tabular}{|c|c|c|c|c|c|c|c|}
\hline
                                    & \multicolumn{3}{c|}{\textbf{$\beta$ = 1}}                    &                                                            & \multicolumn{3}{c|}{\textbf{$\alpha$ = 1}}                                   \\ \hline
\textbf{$\sigma$\textbackslash{}$\alpha$} & \textbf{0.25}                 & \textbf{0.5} & \textbf{2} & \textbf{$\sigma$\textbackslash{}$\beta$} & \textbf{0}                 & \textbf{0.5} & \textbf{1}                    \\ \hline\hline
\textbf{0.1}                        & 45.5                          & 48.03        & 48.33      & \textbf{0.1}                                               & 40.56                      & 45.42        & 49.98 \\ \hline
\textbf{0.2}                        & 40.55                         & 44.88        & 46.12      & \textbf{0.2}                                               & 43.13                      & 47.37        & 49.98                         \\ \hline
\textbf{0.4}                        & 34.75 & 40.19        & 40.88      & \textbf{0.4}                                               & 49.94 & 49.11        & 49.98                         \\ \hline
\end{tabular}
\caption{PSNR values comparing the original CPU implementation with the FPGA implementation for the flower image.}
\label{tab:psnr}
\end{table}
The first half of Table~\ref{tab:psnr} shows us the PSNR values when we perform detail enhancing and detail smoothing operations on the image. In these operations, $\beta = 1$ , which means that only the detail remapping function contributes to the modification of the input image. We see for a fixed value of $\alpha$, as the value of $\sigma$ increases there is a drop in the PSNR value. This is because as the value of $\sigma$ increases, more pixels in the image, possibly including edge pixels, are treated as details by the hardware and modified by the detail remapping function. Therefore a drop in PSNR is expected.

The second half of Table~\ref{tab:psnr} shows us the PSNR values when we perform tone mapping and inverse tone mapping on the image. In these operations, $\alpha = 1$, which means that only the edge remapping function contributes to the modification of the input image. In this case, we see that when the value of $\sigma$ is small, many pixels could be misclassified as edges and the edge remapping function may be wrongly applied, leading to low PSNR values. However, with the increase in the value of $\sigma$, the misclassification rate decreases which leads to the increase in the PSNR value. Overall, it can noted that our approximation scheme to make the Local Laplacian Filtering computationally tractable does not lead to a loss in the image quality measured using either the PSNR metric or by visual inspection.

\subsection{Active vs Inactive Cycles}
We test the efficiency of our hardware pipeline, on a per LPU basis, by counting the number of active and inactive cycles out of the total execution cycles. An execution cycle is considered to be active for an LPU if it is busy processing data and considered inactive if it is waiting for the data. For all the 3 LPUs, we find that the execution efficiency increases with the increase in the PCIe bandwidth. For example, the execution efficiency of LPU $\Tilde{L}_3$ at PCIe bandwidths 32 and 256 bits per clock cycle is 44.9 percent and 92.3 percent respectively.


 
\begin{figure}[t]
\begin{tabular}{cc}
\includegraphics[scale=0.5]{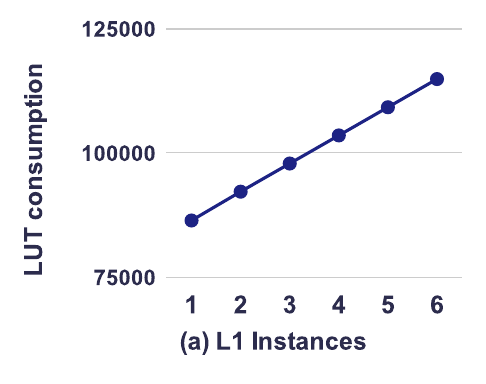} &
\includegraphics[scale=0.5]{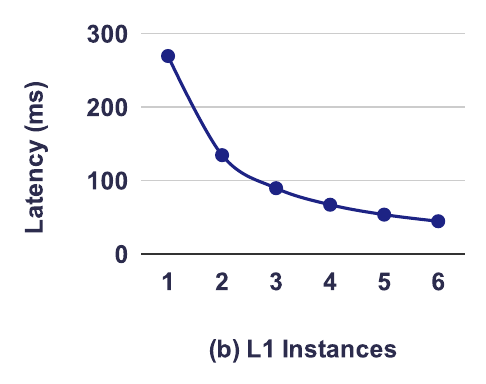}
\end{tabular}
\caption{Graph (a) shows the latency of $\Tilde{L}_1$ with the increase in the number of instances. Graph (b) shows the resource utilization of the hardware with the increase in the number of instances of $\Tilde{L}_1.$}
\label{fig:dse}
\end{figure}

\subsection{Design Space Exploration}
When all the three RGB channels i.e., all the nine processing units as shown in Figure~\ref{fig:cpufpga} run on the Virtex-7 FPGA, the LUT resource consumption is 19\% of the total available on Virtex-7. We know that the computation of each pixel of the output Laplacian pyramid is independent of one another. This leaves us with a lot of scope for hardware scaling by replicating the LPUs and distributing the computations among them to improve the overall latency. Out of the three levels, $\Tilde{L}_1$ computes the maximum number of pixels of the output Laplacian pyramid. $\Tilde{L}_1$ takes the maximum amount of time to complete but has the least resource consumption among $\Tilde{L}_1$, $\Tilde{L}_2$ and $\Tilde{L}_3$ with 1.28\% LUT usage. Therefore, we can scale $\Tilde{L}_1$ by replicating it multiple times in the hardware, thereby dividing the processing among the replicated units to improve overall latency.\par
Graph (a) in Figure~\ref{fig:dse} shows the LUT usage of the hardware as the number of instances increase from of $\Tilde{L}_1$ from 1 to 6. Graph (b) in Figure~\ref{fig:dse} shows that as the number of instances of $\Tilde{L}_1$ increases from 1 to 6 its overall latency reduces drastically showing that the hardware scaling is effective.

%% file: conclusions.tex
\section{Conclusions}\label{sec:conclusions}
In this paper, we proposed a novel pipelined and data parallel architecture for Local Lapalcian Filtering algorithm which is an edge-aware image processing technique. We exploit most of the data parallelism provided by the algorithm. 
In the current architecture, although it is possible to construct intermediate Laplacian pyramids corresponding to different pixels within the same level, we process them one after another. They could be processed in parallel if enough memory bandwidth is available.
Further, we proposed an approximation to the Gaussian filter which enabled us to design a high throughput convolution engine which has both pipeline and data parallelism in it. The Gaussian filter approximation allows for computation of convolutions without using the FPGA DSP blocks. We also found that the LUT usage of our implementation is very low. We believe that because of this low resource utilization our implementation can effectively lend itself to further hardware scaling and can be implemented on lower end boards like ZedBoard.